\documentclass[%
 aip,
 jmp,%
 amsmath,amssymb,
 reprint,%
]{revtex4-1}

\usepackage{subfigure}
\usepackage{graphicx}
\graphicspath{{./}{./images/}{./images_all/}{./images_all/Bennett_clocking_thermal/}}

\usepackage{dcolumn}
\usepackage{bm}

\begin{document}

\preprint{AIP/123-QED}

\title{Critical analysis and remedy of switching failures in straintronic logic using Bennett clocking in the presence of thermal fluctuations}

\author{Kuntal Roy}
\email{royk@purdue.edu.}
\noaffiliation
\affiliation{School of Electrical and Computer Engineering, Purdue University, West Lafayette, Indiana 47907, USA}
\thanks{Work for this paper was performed prior to K. Roy joining Purdue University. Some affiliated work was performed at the School of Electrical and Computer Engineering, Virginia Commonwealth University, Richmond,  Virginia 23284, USA.}


\begin{abstract}
Straintronic logic is a promising platform for beyond Moore's law computing. Using Bennett clocking mechanism, information can propagate through an array of strain-mediated multiferroic nanomagnets exploiting the dipolar coupling between the magnets without having to physically interconnect them. Here we perform a critical analysis of switching failures, i.e., error in information propagation due to thermal fluctuations through a chain of such straintronic devices. We solved stochastic Landau-Lifshitz-Gilbert equation considering room-temperature thermal perturbations and show that magnetization switching may fail due to inherent magnetization dynamics accompanied by thermally broadened switching delay distribution. Avenues available to circumvent such issue are proposed.
\end{abstract}

\maketitle

\begin{figure*}
\centering
\includegraphics{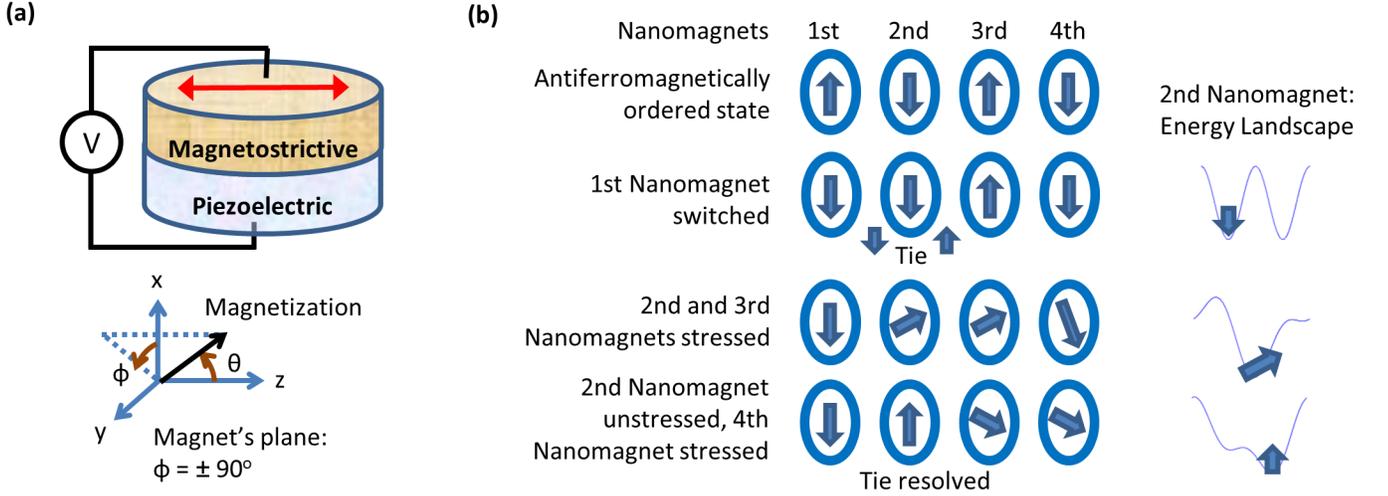}
\caption{\label{fig:Bennett_clocking_introduction} Bennett clocking mechanism for unidirectional information propagation in straintronic logic.
(a) A voltage-controlled strain-mediated multiferroic device and axis assignment. Magnetization is bistable along the $\pm z$-axis, which stores a bit of information $0$ or $1$. 
(b) Unidirectional information propagation through a horizontal chain of straintronic devices. The nanomagnets are stressed separately using different voltage sources. Note that the dipolar coupling between the neighboring nanomagnets is bidirectional and hence we need to impose the unidirectionality in time (using a 3-phase clocking scheme to apply stress on the nanomagnets subsequently) to propagate a bit of information through the chain. The magnetization of the 1st nanomagnet is flipped, and the 2nd and 3rd nanomagnets are stressed to align their magnetizations along the hard-axis. Then stress is released/reversed on the 2nd nanomagnet to relax its magnetization along its desired state.}
\end{figure*}

Multiferroic devices,~\cite{roy11} consisting of a piezoelectric layer strain-coupled to a magnetostrictive nanomagnet, hold profound promise to replace traditional transistors for our future information processing paradigm. These devices work according to the principle of converse magnetoelectric effect,~\cite{RefWorks:558,Refworks:164,Refworks:165} i.e., when a voltage is applied across the device, the piezoelectric layer gets strained and the strain is elastically transferred to the magnetostrictive layer rotating its magnetization (see Fig.~\ref{fig:Bennett_clocking_introduction}a). With appropriate choice of materials, such devices dissipate a minuscule amount of energy of $\sim$1 attojoule in sub-nanosecond switching delay at room-temperature.~\cite{roy11_6} This study has opened up a field called \emph{straintronics}~\cite{roy13_spin, roy13} and experimental efforts to demonstrate such electric-field induced magnetization switching are considerably emerging.~\cite{RefWorks:551,RefWorks:609,RefWorks:611,RefWorks:609}

Information processing using Bennett clocking mechanism~\cite{RefWorks:144} is an attractive platform for building logic using nanomagnets and dipolar coupling between them. This facilitates avoiding physical interconnects and thus eliminating the energy dissipation due to charging and discharging of interconnect capacitances. Figure~\ref{fig:Bennett_clocking_introduction}b depicts how a bit of information can be propagated unidirectionally through a chain of \emph{energy-efficient} stress-mediated multiferroic devices~\cite{RefWorks:154,roy13_spin,fasha11,*fasha11_erratum} rather than using highly energy consuming magnetic field.~\cite{RefWorks:135,RefWorks:663} Contrary to the steady-state analysis,~\cite{RefWorks:154} the investigation of magnetization dynamics~\cite{roy11,roy11_6,roy13_spin} has proved to be crucial for achieving sub-nanosecond switching speed making the straintronic logic competitive with traditional charge-based computing.~\cite{roy13_spin,nano_edi} Recent experiments have demonstrated defects and errors in nanomagnetic logic circuits.~\cite{RefWorks:662}

Here, we study the source of switching failures in straintronic logic due to thermal fluctuations during the propagation of a bit of information. We solved stochastic Landau-Lifshitz-Gilbert (LLG) equation~\cite{RefWorks:162,RefWorks:161,RefWorks:186,roy11_6} of magnetization dynamics to understand the critical issues behind the switching failures. As such we would assume that the magnetization of the 2nd nanomagnet in Fig.~\ref{fig:Bennett_clocking_introduction}b would always switch successfully to the desired state as shown in the last row due to the dipole coupling from the 1st nanomagnet. However, the analysis presented here demonstrates that magnetization's slight excursion out of magnet's plane accompanied by the thermal fluctuations can eventually make magnetization backtracking to the wrong direction. This would produce error in propagating a bit of information. Making an approximation by \emph{not} taking into account the out-of-plane excursion of magnetization~\cite{roy11_2,roy13_2} would not be able to comprehend such critical reasoning behind switching failures. Noting that it requires a very small bit error rate ($< 10^{-4}$) for computing purposes, we further suggest an way to tackle such issue.

We model the magnetostrictive nanomagnet in the shape of an elliptical cylinder; its cross-section lies on the $y$-$z$ plane, the major axis points along the $z$-direction, and the minor axis along the $y$-direction (see Fig.~\ref{fig:Bennett_clocking_introduction}a). Any deflection of magnetization out of magnet's plane ($y$-$z$ plane, $\phi=\pm90^\circ$) is termed as out-of-plane excursion. The dimensions of the major axis, the minor axis, and the thickness are $a$, $b$, and $l$, respectively ($a > b > l$). So the volume is $\Omega=(\pi/4)abl$. We will consider the switching of nanomagnet-2 and the subscript of any parameter will point to the corresponding nanomagnet (1 to 4, see Fig.~\ref{fig:Bennett_clocking_introduction}b).

We solve the stochastic Landau-Lifshitz-Gilbert (LLG) equation~\cite{RefWorks:162,RefWorks:161,RefWorks:186} in the presence of thermal fluctuations (details are provided in the supplementary material~\cite{supplx_deriv}) and derive the following coupled equations for the dynamics of $\theta_2$ and $\phi_2$: 
\begin{multline}
\left(1+\alpha^2 \right) \cfrac{d\theta_2}{dt} = \cfrac{|\gamma|}{M_V} [ B_{shape,\phi_2}(\phi_2)sin\theta_2 \displaybreak[3]\\
 - 2\alpha B_2(\phi_2) sin\theta_2 cos\theta_2 - T_{dipole,\theta_2} - \alpha T_{dipole,\phi_2} \displaybreak[3] \\
 + (\alpha P_{\theta_2} + P_{\phi_2})],
 \label{eq:theta_dynamics_bennett}
\end{multline}
\begin{multline}
\left(1+\alpha^2 \right) \cfrac{d\phi_2}{dt} = \cfrac{|\gamma|}{M_V} \cfrac{1}{sin\theta_2} [\alpha B_{shape,\phi_2}(\phi_2)sin\theta_2 \\
	 + 2 B_2(\phi_2) sin\theta_2 cos\theta_2 + \alpha T_{dipole,\theta_2} + T_{dipole,\phi_2} \\
	 - \{sin\theta_2\}^{-1} (P_{\theta_2} - \alpha P_{\phi_2})] \qquad (sin\theta_2 \neq 0),
  \label{eq:phi_dynamics_bennett}
\end{multline}
where
\begin{equation}
B_{shape,\phi_2}(\phi_2) = (\mu_0/2) \, M_s^2 \Omega (N_{d-xx}-N_{d-yy}) sin(2\phi_2),
\label{eq:B_shape_phi}
\end{equation}
\begin{subequations}
\begin{align}
B_2(\phi_2) &= B_{shape,2}(\phi_2) + B_{stress,2},\displaybreak[3]\\
B_{shape,2}(\phi_2) &= (\mu_0/2) M_s^2 \Omega [(N_{d-yy}-N_{d-zz}) \nonumber \\ 
								& \qquad \qquad + (N_{d-xx}-N_{d-yy})\,cos^2\phi_2],\\
B_{stress,2} 	&= (3/2) \lambda_s \sigma_2 \Omega,
\label{eq:shape_stress}
\end{align}
\end{subequations}
\begin{subequations}
\begin{align}
P_{\theta_2} &= M_V [ h_{x,2}\,cos\theta_2\,cos\phi_2 + h_{y,2}\,cos\theta_2 sin\phi_2  \nonumber \displaybreak[3]\\
						 & \qquad \qquad- h_{z,2}\,sin\theta_2], \displaybreak[3]\\
P_{\phi_2} &= M_V [h_{y,2}\,cos\phi_2 -h_{x,2}\,sin\phi_2], \displaybreak[3]\\
h_{i,2} &= \sqrt{\frac{2 \alpha kT}{|\gamma| M_V \Delta t}} \;G_{(0,1)} \;\;(i=x,y,z), \label{eq:thermal_h}
\end{align}
\end{subequations}
$\alpha$ is the phenomenological damping parameter, $\gamma$ is the gyromagnetic ratio for electrons, $M_V= \mu_0 M_s \Omega$, $M_s$ is the saturation magnetization, $N_{d-mm}$ is the component of demagnetization factor along $m$-direction, which depends on the nanomagnet's dimensions,~\cite{RefWorks:157,RefWorks:402} $(3/2)\lambda_s$ is the magnetostrictive coefficient of the single-domain magnetostrictive nanomagnet,~\cite{RefWorks:157} $\sigma_2$ is the stress on the nanomagnet-2 (note that the product of magnetostrictive coefficient and stress needs to be \emph{negative} in sign for stress-anisotropy to overcome the shape-anisotropy), $\Delta t$ is the simulation time-step, $G_{(0,1)}$ is a Gaussian distribution with zero mean and unit variance,~\cite{RefWorks:388} $k$ is the Boltzmann constant, $T$ is temperature, $T_{dipole,\theta_2} = (1/sin \theta_2)(\partial E_{dipole,2}/\partial \phi_2)$, $T_{dipole,\phi_2} = \partial E_{dipole,2}/\partial \theta_2$, and $E_{dipole,2}$ is the dipole coupling energy from the neighboring nanomagnets 1 and 3. Note that in a very similar way the equations of dynamics for the other three nanomagnets can be derived.

\begin{figure}
\centering
\includegraphics{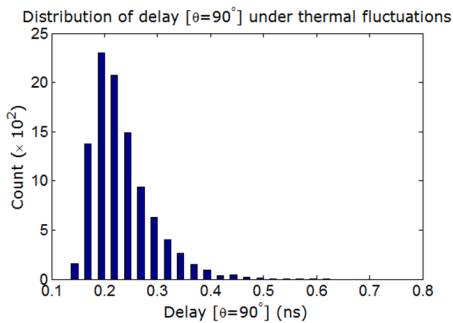}
\caption{\label{fig:Bennett_clocking_distribution_theta_90deg} Distribution of switching delay when magnetization of nanomagnet-2 reaches at $\theta=90^\circ$ from $\theta\simeq180^\circ$ upon application of 20 MPa stress at 100 ps ramp period. A moderately large number (10,000) of simulations have been performed in the presence of room-temperature (300 K) thermal fluctuations to generate this distribution. This wide distribution is caused by the following two reasons: (1) thermal fluctuations make the initial orientation of magnetization a distribution, and (2) thermal kicks during the transition from $\theta\simeq180^\circ$ to $\theta = 90^\circ$ make the time-period a distribution too. The mean and standard deviation of this distribution are 0.232 ns and 0.056 ns, respectively.} 
\end{figure}

The magnetostrictive layer is considered to be made of polycrystalline Terfenol-D, which has the following material properties -- Young's modulus (Y): 80 GPa, saturation magnetization ($M_s$):  8$\times$10$^5$ A/m, Gilbert's damping constant ($\alpha$): 0.1, and magnetostrictive coefficient ($(3/2)\lambda_s$): +90$\times$10$^{-5}$ (Refs.~\onlinecite{RefWorks:179,RefWorks:176,RefWorks:178,roy11_6}). The dimensions of the nanomagnet are chosen as $a=100\,nm$, $b=90\,nm$, and $l=6\,nm$, ensuring the validity of single-domain assumption.~\cite{RefWorks:402,RefWorks:133} The center-to-center distance between the nanomagnets is chosen as $R=120\,nm$. 

The piezoelectric layer is made of PMN-PT,~\cite{pmnpt} which has a dielectric constant of 1000 and the layer is assumed to be four times thicker than the magnetostrictive layer.~\cite{roy11} Assuming that maximum strain that can be generated in the piezoelectric layer is 500 ppm,~\cite{RefWorks:170,RefWorks:563} it would require an electric field of $\sim$0.4 MV/m because $d_{31}$=13$\times$10$^{-10}$ m/V for PMN-PT.~\cite{pmnpt} The stress generated in the Terfenol-D layer is the product of strain and Young's modulus. Hence, 4.6 mVs of voltages would generate 20 MPa stress in the Terfenol-D layer. 

\begin{figure*}
\centering
\includegraphics{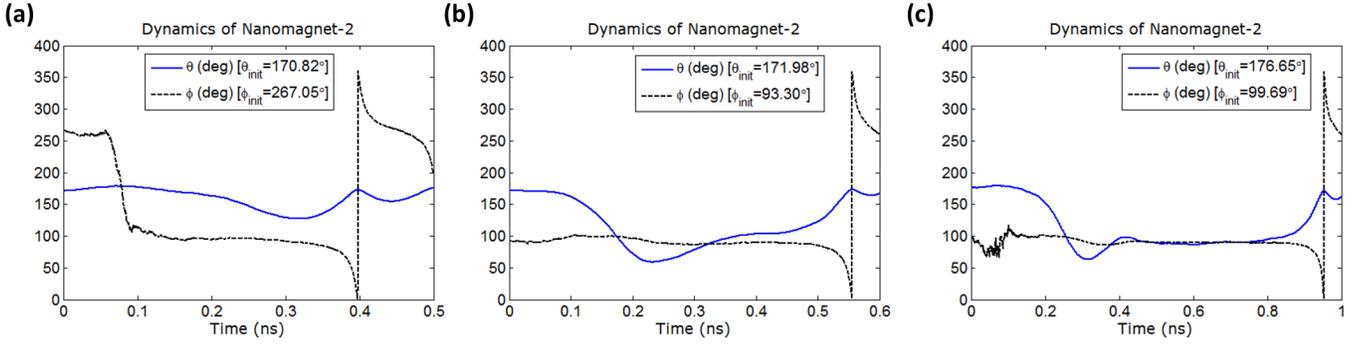}
\caption{\label{fig:Bennett_clocking_switching_failures} Magnetization of nanomagnet-2 backtracks to the same easy axis it started resulting in an error in propagating a bit of information. 
(a) Stress is ramped up from zero to 20 MPa in 100 ps, kept constant for 100 ps, and ramped down (and reversed) at the same rate as for ramp-up. Magnetization failed to even reach at $\theta=90^\circ$ since stress was not active for a sufficient amount of time. Note the $\phi$-dynamics that magnetization started close from an in-plane angle $\phi=270^\circ (-90^\circ)$, and traversed to another in-plane angle $\phi=90^\circ$ due to $\theta$-$\phi$ coupled dynamics.
(b) Stress is kept constant longer (for 200 ps) and ramp up/down times are same as for the part (a). Magnetization switching failure is still observed. However, this time magnetization was able to get past $\theta=90^\circ$ towards $\theta=0^\circ$, but during the ramp-down phase, magnetization was subjected to a detrimental motion forcing magnetization to backtrack towards $\theta=180^\circ$.
(c) Stress is kept constant even longer (for 600 ps) and ramp up/down times are same as for the part (a). Magnetization switching failure is still observed. Magnetization is lingering around $\theta=90^\circ$ since stress is kept constant for long time, however, as stress is brought down, magnetization started backtracking towards $\theta=180^\circ$.
}
\end{figure*}

Figure~\ref{fig:Bennett_clocking_distribution_theta_90deg} shows that upon application of stress, different trajectories of magnetization of nanomagnet-2 reaches $\theta=90^\circ$ at variable times in the presence of thermal fluctuations. We take the distribution of initial orientation of magnetization due to thermal fluctuations into account. If we release/reverse the stress on nanomagnet-2 ahead of time, then magnetization may not be able to switch successfully rather it will backtrack to the same easy axis it started. This is exemplified in Fig.~\ref{fig:Bennett_clocking_switching_failures}a. Magnetization failed to switch to $\theta \simeq 0^\circ$, even it could not get past $\theta \simeq 90^\circ$ since stress is kept constant for a short period of time of 100 ps. Out of 10,000 simulations, 16.52\% switching failures were observed in this case. If we keep the stress constant longer for 200 ps, it would not necessarily result in reducing the failure rate of switching, which we will discuss later. Figure~\ref{fig:Bennett_clocking_switching_failures}b depicts a case when switching failed; magnetization backtracked when stress was ramped down. We keep the stress constant for much longer time (600 ps) to observe whether sometimes switching still fails or not; Fig.~\ref{fig:Bennett_clocking_switching_failures}c shows a case when switching fails. Note that magnetization keeps lingering around $\theta=90^\circ$ since stress was kept constant for long time, but magnetization eventually backtracked towards $\theta=180^\circ$.


\begin{figure*}
\centering
\includegraphics{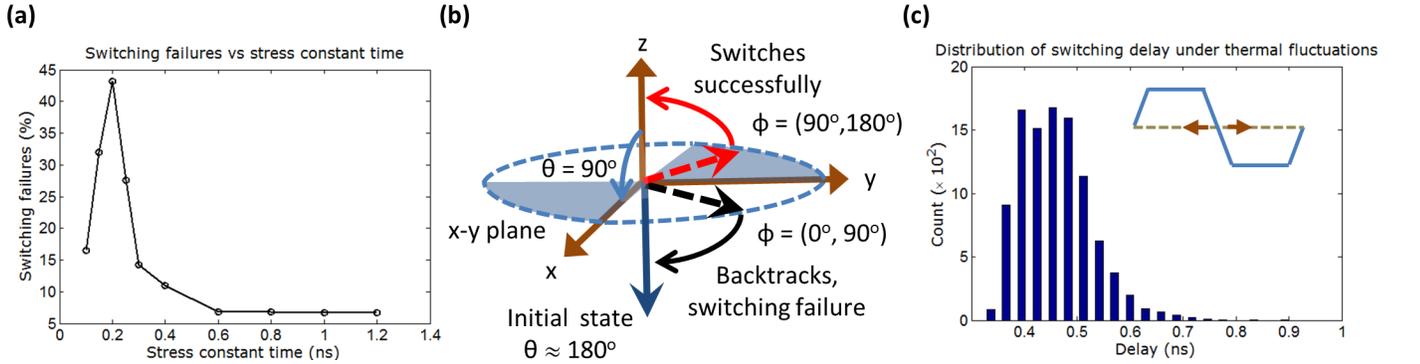}
\caption{\label{fig:Bennett_clocking_switching_failures_analysis} 
(a) Percentage switching failures versus stress constant time for 20 MPa stress and 100 ps ramp period. Failure rate increases with stress constant time initially but later decreases with further increasing stress constant time before getting saturated.
(b) Out-of-plane ($\phi \neq \pm 90^\circ$) excursion of magnetization during its dynamical motion. The magnetization deflects out-of-plane due to the torque exerted on it and fast (not adiabatic) ramp rate. When magnetization reaches around $\theta=90^\circ$ ($x$-$y$ plane), if it resides in the quadrant ($90^\circ$,$180^\circ$) [or ($270^\circ$,$360^\circ$)], it will aid magnetization's motion towards $\theta=0^\circ$; otherwise, residing in the other two quadrants would be tantamount to switching failure. 
(c) Stress is released dynamically when magnetization reaches $\theta=90^\circ$. Magnetization always switches successfully in this methodology. The mean and standard deviation of this distribution are 0.459 ns and 0.067 ns, respectively.
}
\end{figure*}

Figure~\ref{fig:Bennett_clocking_switching_failures_analysis}a depicts the non-trivial dependence of switching failure rate with stress constant time. This can be explained by considering magnetization's excursion out of magnet's plane ($y$-$z$ plane, $\phi=\pm90^\circ$) during switching as shown in Fig.~\ref{fig:Bennett_clocking_switching_failures_analysis}b. Magnetization deflects out-of-plane due to the torque exerted on it in the $\mathbf{\hat{e}_\phi}$ direction and fast (non-adiabatic) ramp of stress. Although the dipole coupling from nanomagnet-1 facilitates switching of magnetization of nanomagnet-2 towards $\theta=0^\circ$, the out-of-plane excursion of nanomagnet-2's magnetization can hinder the switching. In Fig.~\ref{fig:Bennett_clocking_switching_failures_analysis}b, note that if magnetization resides in the quadrant $\phi \in (90^\circ, 180^\circ)$ [or $\phi \in (270^\circ, 360^\circ)$], the term $B_{shape,\phi_2}(\phi_2)$ as in the Eq.~\eqref{eq:B_shape_phi} becomes negative in sign, which would facilitate decreasing the value of $\theta_2$ [see Eq.~\eqref{eq:theta_dynamics_bennett}] aiding magnetization rotation in the correct direction. But, if magnetization resides in the other two quadrants [$\phi \in (90^\circ, 180^\circ)$ or $\phi \in (270^\circ, 360^\circ)$, termed as \emph{bad} quadrants onwards], the term $B_{shape,\phi_2}(\phi_2)$ would be positive, which would force magnetization backtracking towards $\theta=180^\circ$. This inherent motion is generated particularly due to $\phi$-dependence of potential energy, which is strong enough to affect the magnetization dynamics. This motion is also responsible for reducing the switching delay by a couple of orders in magnitude and bellied under the fact that $N_{d-xx} \gg N_{d-yy}, N_{d-zz}$ ($N_{d-xx}-N_{d-yy}$ is higher than $N_{d-yy}-N_{d-zz}$ by a couple of orders in magnitude).

When magnetization reaches $\theta=90^\circ$, thermal fluctuations can scuttle magnetization in either side of the magnet's plane ($\phi=\pm90^\circ$). As explained earlier, if magnetization happens to be in the \emph{bad} quadrants, magnetization would be prone to backtracking, causing a switching failure. In Fig.~\ref{fig:Bennett_clocking_switching_failures_analysis}a, when the stress constant time is low, any increase in stress constant time has two effects that counter each other: (1) More trajectories finish before stress is released (see Fig.~\ref{fig:Bennett_clocking_distribution_theta_90deg}) so magnetization for those trajectories keep lingering around $\theta=90^\circ$ and thermal fluctuations may cause mishap by scuttling magnetization in the \emph{bad} quadrants for $\phi$. This will increase the switching failure rate. (2) More trajectories will be able to reach at $\theta=90^\circ$ before we release the stress (see Fig.~\ref{fig:Bennett_clocking_distribution_theta_90deg}), whereupon magnetization can possibly switch successfully towards $\theta=0^\circ$. This will decrease the switching failure rate. Due to these two counteracting effects, we do see a peak in the Fig.~\ref{fig:Bennett_clocking_switching_failures_analysis}a. Now, if we keep increasing the stress constant time so that all the trajectories reach at $\theta=90^\circ$ before stress is released, the switching failure rate saturates to $\sim$7\%. The reason is that thermal fluctuations scuttle magnetization in the bad quadrants for $\phi$ upon reaching $\theta=90^\circ$ and dipole coupling from nanomagnet-1 cannot help much. This saturated switching failure rate of $\sim$7\% can be decreased a bit by optimizing parameters but it is not possible to decrease it to a very small value ($ < 10^{-4}$) for general-purpose computing purposes. Also, waiting until all trajectories reach at $\theta=90^\circ$, increases the switching time period. Hence, we propose the following approach.

We realize that letting magnetization collapse on the magnet's plane and thermal fluctuations scuttling magnetization in the \emph{bad} quadrants are causing the switching failures. Provided we apply a sufficiently high stress with a sufficiently fast ramp rate, magnetization will not traverse into \emph{bad} quadrants during switching while magnetization passes through the $x$-$y$ plane.~\cite{roy13_2} We can use a sensing circuitry to detect when magnetization reaches around $\theta=90^\circ$ so that we can ramp down the stress thereafter. The sensing circuitry can be implemented by measuring the magnetoresistance in a magnetic tunnel junction (MTJ).~\cite{RefWorks:577,RefWorks:555,RefWorks:572,RefWorks:76,RefWorks:74,RefWorks:33,RefWorks:300,roy13_2} Basically, we need to get calibrated on the magnetoresistance of the MTJ when magnetization resides on the $x$-$y$ plane. Comparing this known signal with the sensed signal of the MTJ, the stress can be ramped down. Figure~\ref{fig:Bennett_clocking_switching_failures_analysis}c shows the distribution of switching delay considering such sensing circuitry. No switching failures were observed and the mean energy dissipation in the nanomagnets turns out to be $\sim$1.5 attojoules at sub-nanosecond switching delay. Some tolerance is nonetheless required since the sensing circuitry cannot be perfect. We performed simulations to show that internal dynamics works correctly as long as the stress is ramped down when magnetization's orientation is in the interval $\theta \in (85^\circ, 140^\circ)$, i.e. it does not have to be exactly $90^\circ$. This tolerance is due to the motion arising from the out-of-plane excursion of magnetization. 

In conclusion, we have performed a critical analysis of switching failures in energy-efficient straintronic logic using Bennett clocking for computing purposes. It is shown that the switching failures are caused by the inherent magnetization dynamics particularly due to out-of-plane excursion of magnetization and thermal fluctuations during switching. We have proposed a novel approach to circumvent such basic issue after a thorough analysis. Such methodology can be exploited for building logic gates and general-purpose computing purposes. Bennett clocking based architecture is regular in nature, so that different building blocks for computing purposes can be designed systematically. Such energy-efficient, fast, and non-volatile (that can lead to instant turn-on computer) computing methodology has profound promise of being the staple of our future information processing paradigm. Processors based on this paradigm may be suitable for applications that need to be run from energy harvested from the environment e.g., wireless sensor networks, medically implanted devices monitoring epileptic patient's brain to warn an impending seizure.


\begin{thebibliography}{41}%
\makeatletter
\providecommand \@ifxundefined [1]{%
 \@ifx{#1\undefined}
}%
\providecommand \@ifnum [1]{%
 \ifnum #1\expandafter \@firstoftwo
 \else \expandafter \@secondoftwo
 \fi
}%
\providecommand \@ifx [1]{%
 \ifx #1\expandafter \@firstoftwo
 \else \expandafter \@secondoftwo
 \fi
}%
\providecommand \natexlab [1]{#1}%
\providecommand \enquote  [1]{``#1''}%
\providecommand \bibnamefont  [1]{#1}%
\providecommand \bibfnamefont [1]{#1}%
\providecommand \citenamefont [1]{#1}%
\providecommand \href@noop [0]{\@secondoftwo}%
\providecommand \href [0]{\begingroup \@sanitize@url \@href}%
\providecommand \@href[1]{\@@startlink{#1}\@@href}%
\providecommand \@@href[1]{\endgroup#1\@@endlink}%
\providecommand \@sanitize@url [0]{\catcode `\\12\catcode `\$12\catcode
  `\&12\catcode `\#12\catcode `\^12\catcode `\_12\catcode `\%12\relax}%
\providecommand \@@startlink[1]{}%
\providecommand \@@endlink[0]{}%
\providecommand \url  [0]{\begingroup\@sanitize@url \@url }%
\providecommand \@url [1]{\endgroup\@href {#1}{\urlprefix }}%
\providecommand \urlprefix  [0]{URL }%
\providecommand \Eprint [0]{\href }%
\providecommand \doibase [0]{http://dx.doi.org/}%
\providecommand \selectlanguage [0]{\@gobble}%
\providecommand \bibinfo  [0]{\@secondoftwo}%
\providecommand \bibfield  [0]{\@secondoftwo}%
\providecommand \translation [1]{[#1]}%
\providecommand \BibitemOpen [0]{}%
\providecommand \bibitemStop [0]{}%
\providecommand \bibitemNoStop [0]{.\EOS\space}%
\providecommand \EOS [0]{\spacefactor3000\relax}%
\providecommand \BibitemShut  [1]{\csname bibitem#1\endcsname}%
\let\auto@bib@innerbib\@empty
\bibitem [{\citenamefont {Roy}, \citenamefont {Bandyopadhyay},\ and\
  \citenamefont {Atulasimha}(2011{\natexlab{a}})}]{roy11}%
  \BibitemOpen
  \bibfield  {author} {\bibinfo {author} {\bibfnamefont {K.}~\bibnamefont
  {Roy}}, \bibinfo {author} {\bibfnamefont {S.}~\bibnamefont {Bandyopadhyay}},
  \ and\ \bibinfo {author} {\bibfnamefont {J.}~\bibnamefont {Atulasimha}},\
  }\href@noop {} {\bibfield  {journal} {\bibinfo  {journal} {Appl. Phys.
  Lett.}\ }\textbf {\bibinfo {volume} {99}},\ \bibinfo {pages} {063108}
  (\bibinfo {year} {2011}{\natexlab{a}})}\BibitemShut {NoStop}%
\bibitem [{\citenamefont {Spaldin}\ and\ \citenamefont
  {Fiebig}(2005)}]{RefWorks:558}%
  \BibitemOpen
  \bibfield  {author} {\bibinfo {author} {\bibfnamefont {N.~A.}\ \bibnamefont
  {Spaldin}}\ and\ \bibinfo {author} {\bibfnamefont {M.}~\bibnamefont
  {Fiebig}},\ }\href@noop {} {\bibfield  {journal} {\bibinfo  {journal}
  {Science}\ }\textbf {\bibinfo {volume} {309}},\ \bibinfo {pages} {391}
  (\bibinfo {year} {2005})}\BibitemShut {NoStop}%
\bibitem [{\citenamefont {Eerenstein}, \citenamefont {Mathur},\ and\
  \citenamefont {Scott}(2006)}]{Refworks:164}%
  \BibitemOpen
  \bibfield  {author} {\bibinfo {author} {\bibfnamefont {W.}~\bibnamefont
  {Eerenstein}}, \bibinfo {author} {\bibfnamefont {N.~D.}\ \bibnamefont
  {Mathur}}, \ and\ \bibinfo {author} {\bibfnamefont {J.~F.}\ \bibnamefont
  {Scott}},\ }\href@noop {} {\bibfield  {journal} {\bibinfo  {journal}
  {Nature}\ }\textbf {\bibinfo {volume} {442}},\ \bibinfo {pages} {759}
  (\bibinfo {year} {2006})}\BibitemShut {NoStop}%
\bibitem [{\citenamefont {Nan}\ \emph {et~al.}(2008)\citenamefont {Nan},
  \citenamefont {Bichurin}, \citenamefont {Dong}, \citenamefont {Viehland},\
  and\ \citenamefont {Srinivasan}}]{Refworks:165}%
  \BibitemOpen
  \bibfield  {author} {\bibinfo {author} {\bibfnamefont {C.~W.}\ \bibnamefont
  {Nan}}, \bibinfo {author} {\bibfnamefont {M.~I.}\ \bibnamefont {Bichurin}},
  \bibinfo {author} {\bibfnamefont {S.}~\bibnamefont {Dong}}, \bibinfo {author}
  {\bibfnamefont {D.}~\bibnamefont {Viehland}}, \ and\ \bibinfo {author}
  {\bibfnamefont {G.}~\bibnamefont {Srinivasan}},\ }\href@noop {} {\bibfield
  {journal} {\bibinfo  {journal} {J. Appl. Phys.}\ }\textbf {\bibinfo {volume}
  {103}},\ \bibinfo {pages} {031101} (\bibinfo {year} {2008})}\BibitemShut
  {NoStop}%
\bibitem [{\citenamefont {Roy}, \citenamefont {Bandyopadhyay},\ and\
  \citenamefont {Atulasimha}(2012)}]{roy11_6}%
  \BibitemOpen
  \bibfield  {author} {\bibinfo {author} {\bibfnamefont {K.}~\bibnamefont
  {Roy}}, \bibinfo {author} {\bibfnamefont {S.}~\bibnamefont {Bandyopadhyay}},
  \ and\ \bibinfo {author} {\bibfnamefont {J.}~\bibnamefont {Atulasimha}},\
  }\href@noop {} {\bibfield  {journal} {\bibinfo  {journal} {J. Appl. Phys.}\
  }\textbf {\bibinfo {volume} {112}},\ \bibinfo {pages} {023914} (\bibinfo
  {year} {2012})}\BibitemShut {NoStop}%
\bibitem [{\citenamefont {Roy}(2013{\natexlab{a}})}]{roy13_spin}%
  \BibitemOpen
  \bibfield  {author} {\bibinfo {author} {\bibfnamefont {K.}~\bibnamefont
  {Roy}},\ }\href@noop {} {\bibfield  {journal} {\bibinfo  {journal} {SPIN}\
  }\textbf {\bibinfo {volume} {3}},\ \bibinfo {pages} {1330003} (\bibinfo
  {year} {2013}{\natexlab{a}})}\BibitemShut {NoStop}%
\bibitem [{\citenamefont {Roy}(2013{\natexlab{b}})}]{roy13}%
  \BibitemOpen
  \bibfield  {author} {\bibinfo {author} {\bibfnamefont {K.}~\bibnamefont
  {Roy}},\ }\href@noop {} {\bibfield  {journal} {\bibinfo  {journal} {Appl.
  Phys. Lett.}\ }\textbf {\bibinfo {volume} {103}},\ \bibinfo {pages} {173110}
  (\bibinfo {year} {2013}{\natexlab{b}})}\BibitemShut {NoStop}%
\bibitem [{\citenamefont {Wu}\ \emph {et~al.}(2011)\citenamefont {Wu},
  \citenamefont {Bur}, \citenamefont {Wong}, \citenamefont {Zhao},
  \citenamefont {Lynch}, \citenamefont {Amiri}, \citenamefont {Wang},\ and\
  \citenamefont {Carman}}]{RefWorks:551}%
  \BibitemOpen
  \bibfield  {author} {\bibinfo {author} {\bibfnamefont {T.}~\bibnamefont
  {Wu}}, \bibinfo {author} {\bibfnamefont {A.}~\bibnamefont {Bur}}, \bibinfo
  {author} {\bibfnamefont {K.}~\bibnamefont {Wong}}, \bibinfo {author}
  {\bibfnamefont {P.}~\bibnamefont {Zhao}}, \bibinfo {author} {\bibfnamefont
  {C.~S.}\ \bibnamefont {Lynch}}, \bibinfo {author} {\bibfnamefont {P.~K.}\
  \bibnamefont {Amiri}}, \bibinfo {author} {\bibfnamefont {K.~L.}\ \bibnamefont
  {Wang}}, \ and\ \bibinfo {author} {\bibfnamefont {G.~P.}\ \bibnamefont
  {Carman}},\ }\href@noop {} {\bibfield  {journal} {\bibinfo  {journal} {Appl.
  Phys. Lett.}\ }\textbf {\bibinfo {volume} {98}},\ \bibinfo {pages} {262504}
  (\bibinfo {year} {2011})}\BibitemShut {NoStop}%
\bibitem [{\citenamefont {Lei}\ \emph {et~al.}(2013)\citenamefont {Lei},
  \citenamefont {Devolder}, \citenamefont {Agnus}, \citenamefont {Aubert},
  \citenamefont {Daniel}, \citenamefont {Kim}, \citenamefont {Zhao},
  \citenamefont {Trypiniotis}, \citenamefont {Cowburn}, \citenamefont {Daniel},
  \citenamefont {Ravelosona},\ and\ \citenamefont {Lecoeur}}]{RefWorks:609}%
  \BibitemOpen
  \bibfield  {author} {\bibinfo {author} {\bibfnamefont {N.}~\bibnamefont
  {Lei}}, \bibinfo {author} {\bibfnamefont {T.}~\bibnamefont {Devolder}},
  \bibinfo {author} {\bibfnamefont {G.}~\bibnamefont {Agnus}}, \bibinfo
  {author} {\bibfnamefont {P.}~\bibnamefont {Aubert}}, \bibinfo {author}
  {\bibfnamefont {L.}~\bibnamefont {Daniel}}, \bibinfo {author} {\bibfnamefont
  {J.}~\bibnamefont {Kim}}, \bibinfo {author} {\bibfnamefont {W.}~\bibnamefont
  {Zhao}}, \bibinfo {author} {\bibfnamefont {T.}~\bibnamefont {Trypiniotis}},
  \bibinfo {author} {\bibfnamefont {R.~P.}\ \bibnamefont {Cowburn}}, \bibinfo
  {author} {\bibfnamefont {L.}~\bibnamefont {Daniel}}, \bibinfo {author}
  {\bibfnamefont {D.}~\bibnamefont {Ravelosona}}, \ and\ \bibinfo {author}
  {\bibfnamefont {P.}~\bibnamefont {Lecoeur}},\ }\href@noop {} {\bibfield
  {journal} {\bibinfo  {journal} {Nature Commun.}\ }\textbf {\bibinfo {volume}
  {4}},\ \bibinfo {pages} {1378} (\bibinfo {year} {2013})}\BibitemShut
  {NoStop}%
\bibitem [{\citenamefont {Kim}\ \emph {et~al.}(2013)\citenamefont {Kim},
  \citenamefont {Schelhas}, \citenamefont {Keller}, \citenamefont {Hockel},
  \citenamefont {Tolbert},\ and\ \citenamefont {Carman}}]{RefWorks:611}%
  \BibitemOpen
  \bibfield  {author} {\bibinfo {author} {\bibfnamefont {H.~K.~D.}\
  \bibnamefont {Kim}}, \bibinfo {author} {\bibfnamefont {L.~T.}\ \bibnamefont
  {Schelhas}}, \bibinfo {author} {\bibfnamefont {S.}~\bibnamefont {Keller}},
  \bibinfo {author} {\bibfnamefont {J.~L.}\ \bibnamefont {Hockel}}, \bibinfo
  {author} {\bibfnamefont {S.~H.}\ \bibnamefont {Tolbert}}, \ and\ \bibinfo
  {author} {\bibfnamefont {G.~P.}\ \bibnamefont {Carman}},\ }\href@noop {}
  {\bibfield  {journal} {\bibinfo  {journal} {Nano Lett.}\ }\textbf {\bibinfo
  {volume} {13}},\ \bibinfo {pages} {884} (\bibinfo {year} {2013})}\BibitemShut
  {NoStop}%
\bibitem [{\citenamefont {Bennett}(1982)}]{RefWorks:144}%
  \BibitemOpen
  \bibfield  {author} {\bibinfo {author} {\bibfnamefont {C.~H.}\ \bibnamefont
  {Bennett}},\ }\href@noop {} {\bibfield  {journal} {\bibinfo  {journal} {Int.
  J. Theor. Phys.}\ }\textbf {\bibinfo {volume} {21}},\ \bibinfo {pages} {905}
  (\bibinfo {year} {1982})}\BibitemShut {NoStop}%
\bibitem [{\citenamefont {Atulasimha}\ and\ \citenamefont
  {Bandyopadhyay}(2010)}]{RefWorks:154}%
  \BibitemOpen
  \bibfield  {author} {\bibinfo {author} {\bibfnamefont {J.}~\bibnamefont
  {Atulasimha}}\ and\ \bibinfo {author} {\bibfnamefont {S.}~\bibnamefont
  {Bandyopadhyay}},\ }\href@noop {} {\bibfield  {journal} {\bibinfo  {journal}
  {Appl. Phys. Lett.}\ }\textbf {\bibinfo {volume} {97}},\ \bibinfo {pages}
  {173105} (\bibinfo {year} {2010})}\BibitemShut {NoStop}%
\bibitem [{\citenamefont {Fashami}\ \emph
  {et~al.}(2011{\natexlab{a}})\citenamefont {Fashami}, \citenamefont {Roy},
  \citenamefont {Atulasimha},\ and\ \citenamefont {Bandyopadhyay}}]{fasha11}%
  \BibitemOpen
  \bibfield  {author} {\bibinfo {author} {\bibfnamefont {M.~S.}\ \bibnamefont
  {Fashami}}, \bibinfo {author} {\bibfnamefont {K.}~\bibnamefont {Roy}},
  \bibinfo {author} {\bibfnamefont {J.}~\bibnamefont {Atulasimha}}, \ and\
  \bibinfo {author} {\bibfnamefont {S.}~\bibnamefont {Bandyopadhyay}},\
  }\href@noop {} {\bibfield  {journal} {\bibinfo  {journal} {Nanotechnology}\
  }\textbf {\bibinfo {volume} {22}},\ \bibinfo {pages} {155201} (\bibinfo
  {year} {2011}{\natexlab{a}})}\BibitemShut {NoStop}%
\bibitem [{\citenamefont {Fashami}\ \emph
  {et~al.}(2011{\natexlab{b}})\citenamefont {Fashami}, \citenamefont {Roy},
  \citenamefont {Atulasimha},\ and\ \citenamefont
  {Bandyopadhyay}}]{fasha11_erratum}%
  \BibitemOpen
  \bibfield  {author} {\bibinfo {author} {\bibfnamefont {M.~S.}\ \bibnamefont
  {Fashami}}, \bibinfo {author} {\bibfnamefont {K.}~\bibnamefont {Roy}},
  \bibinfo {author} {\bibfnamefont {J.}~\bibnamefont {Atulasimha}}, \ and\
  \bibinfo {author} {\bibfnamefont {S.}~\bibnamefont {Bandyopadhyay}},\
  }\href@noop {} {\bibfield  {journal} {\bibinfo  {journal} {Nanotechnology}\
  }\textbf {\bibinfo {volume} {22}},\ \bibinfo {pages} {309501} (\bibinfo
  {year} {2011}{\natexlab{b}})}\BibitemShut {NoStop}%
\bibitem [{\citenamefont {Csaba}\ \emph {et~al.}(2002)\citenamefont {Csaba},
  \citenamefont {Imre}, \citenamefont {Bernstein}, \citenamefont {Porod},\ and\
  \citenamefont {Metlushko}}]{RefWorks:135}%
  \BibitemOpen
  \bibfield  {author} {\bibinfo {author} {\bibfnamefont {G.}~\bibnamefont
  {Csaba}}, \bibinfo {author} {\bibfnamefont {A.}~\bibnamefont {Imre}},
  \bibinfo {author} {\bibfnamefont {G.~H.}\ \bibnamefont {Bernstein}}, \bibinfo
  {author} {\bibfnamefont {W.}~\bibnamefont {Porod}}, \ and\ \bibinfo {author}
  {\bibfnamefont {V.}~\bibnamefont {Metlushko}},\ }\href@noop {} {\bibfield
  {journal} {\bibinfo  {journal} {IEEE Trans. Nanotech.}\ }\textbf {\bibinfo
  {volume} {1}},\ \bibinfo {pages} {209} (\bibinfo {year} {2002})}\BibitemShut
  {NoStop}%
\bibitem [{\citenamefont {Liu}\ \emph {et~al.}(2013)\citenamefont {Liu},
  \citenamefont {Hu}, \citenamefont {Niemier}, \citenamefont {Nahas},
  \citenamefont {Csaba}, \citenamefont {Bernstein},\ and\ \citenamefont
  {Porod}}]{RefWorks:663}%
  \BibitemOpen
  \bibfield  {author} {\bibinfo {author} {\bibfnamefont {S.}~\bibnamefont
  {Liu}}, \bibinfo {author} {\bibfnamefont {X.}~\bibnamefont {Hu}}, \bibinfo
  {author} {\bibfnamefont {M.~T.}\ \bibnamefont {Niemier}}, \bibinfo {author}
  {\bibfnamefont {J.}~\bibnamefont {Nahas}}, \bibinfo {author} {\bibfnamefont
  {G.}~\bibnamefont {Csaba}}, \bibinfo {author} {\bibfnamefont
  {G.}~\bibnamefont {Bernstein}}, \ and\ \bibinfo {author} {\bibfnamefont
  {W.}~\bibnamefont {Porod}},\ }\href@noop {} {\bibfield  {journal} {\bibinfo
  {journal} {IEEE Trans. Nanotech.}\ }\textbf {\bibinfo {volume} {12}},\
  \bibinfo {pages} {203} (\bibinfo {year} {2013})}\BibitemShut {NoStop}%
\bibitem [{\citenamefont {Demming}(2012)}]{nano_edi}%
  \BibitemOpen
  \bibfield  {author} {\bibinfo {author} {\bibfnamefont {A.}~\bibnamefont
  {Demming}},\ }\href@noop {} {\bibfield  {journal} {\bibinfo  {journal}
  {Nanotechnology}\ }\textbf {\bibinfo {volume} {23}},\ \bibinfo {pages}
  {390201} (\bibinfo {year} {2012})}\BibitemShut {NoStop}%
\bibitem [{\citenamefont {Carlton}\ \emph {et~al.}(2012)\citenamefont
  {Carlton}, \citenamefont {Lambson}, \citenamefont {Scholl}, \citenamefont
  {Young}, \citenamefont {Ashby}, \citenamefont {Dhuey},\ and\ \citenamefont
  {Bokor}}]{RefWorks:662}%
  \BibitemOpen
  \bibfield  {author} {\bibinfo {author} {\bibfnamefont {D.}~\bibnamefont
  {Carlton}}, \bibinfo {author} {\bibfnamefont {B.}~\bibnamefont {Lambson}},
  \bibinfo {author} {\bibfnamefont {A.}~\bibnamefont {Scholl}}, \bibinfo
  {author} {\bibfnamefont {A.}~\bibnamefont {Young}}, \bibinfo {author}
  {\bibfnamefont {P.}~\bibnamefont {Ashby}}, \bibinfo {author} {\bibfnamefont
  {S.}~\bibnamefont {Dhuey}}, \ and\ \bibinfo {author} {\bibfnamefont
  {J.}~\bibnamefont {Bokor}},\ }\href@noop {} {\bibfield  {journal} {\bibinfo
  {journal} {IEEE Trans. Nanotech.}\ }\textbf {\bibinfo {volume} {11}},\
  \bibinfo {pages} {760} (\bibinfo {year} {2012})}\BibitemShut {NoStop}%
\bibitem [{\citenamefont {Landau}\ and\ \citenamefont
  {Lifshitz}(1935)}]{RefWorks:162}%
  \BibitemOpen
  \bibfield  {author} {\bibinfo {author} {\bibfnamefont {L.}~\bibnamefont
  {Landau}}\ and\ \bibinfo {author} {\bibfnamefont {E.}~\bibnamefont
  {Lifshitz}},\ }\href@noop {} {\bibfield  {journal} {\bibinfo  {journal}
  {Phys. Z. Sowjet.}\ }\textbf {\bibinfo {volume} {8}},\ \bibinfo {pages} {101}
  (\bibinfo {year} {1935})}\BibitemShut {NoStop}%
\bibitem [{\citenamefont {Gilbert}(2004)}]{RefWorks:161}%
  \BibitemOpen
  \bibfield  {author} {\bibinfo {author} {\bibfnamefont {T.~L.}\ \bibnamefont
  {Gilbert}},\ }\href@noop {} {\bibfield  {journal} {\bibinfo  {journal} {IEEE
  Trans. Magn.}\ }\textbf {\bibinfo {volume} {40}},\ \bibinfo {pages} {3443}
  (\bibinfo {year} {2004})}\BibitemShut {NoStop}%
\bibitem [{\citenamefont {Brown}(1963)}]{RefWorks:186}%
  \BibitemOpen
  \bibfield  {author} {\bibinfo {author} {\bibfnamefont {W.~F.}\ \bibnamefont
  {Brown}},\ }\href@noop {} {\bibfield  {journal} {\bibinfo  {journal} {Phys.
  Rev.}\ }\textbf {\bibinfo {volume} {130}},\ \bibinfo {pages} {1677} (\bibinfo
  {year} {1963})}\BibitemShut {NoStop}%
\bibitem [{\citenamefont {Roy}, \citenamefont {Bandyopadhyay},\ and\
  \citenamefont {Atulasimha}(2011{\natexlab{b}})}]{roy11_2}%
  \BibitemOpen
  \bibfield  {author} {\bibinfo {author} {\bibfnamefont {K.}~\bibnamefont
  {Roy}}, \bibinfo {author} {\bibfnamefont {S.}~\bibnamefont {Bandyopadhyay}},
  \ and\ \bibinfo {author} {\bibfnamefont {J.}~\bibnamefont {Atulasimha}},\
  }\href@noop {} {\bibfield  {journal} {\bibinfo  {journal} {Phys. Rev. B}\
  }\textbf {\bibinfo {volume} {83}},\ \bibinfo {pages} {224412} (\bibinfo
  {year} {2011}{\natexlab{b}})}\BibitemShut {NoStop}%
\bibitem [{\citenamefont {Roy}, \citenamefont {Bandyopadhyay},\ and\
  \citenamefont {Atulasimha}(2013)}]{roy13_2}%
  \BibitemOpen
  \bibfield  {author} {\bibinfo {author} {\bibfnamefont {K.}~\bibnamefont
  {Roy}}, \bibinfo {author} {\bibfnamefont {S.}~\bibnamefont {Bandyopadhyay}},
  \ and\ \bibinfo {author} {\bibfnamefont {J.}~\bibnamefont {Atulasimha}},\
  }\href@noop {} {\bibfield  {journal} {\bibinfo  {journal} {Sci. Rep.}\
  }\textbf {\bibinfo {volume} {3}},\ \bibinfo {pages} {3038} (\bibinfo {year}
  {2013})}\BibitemShut {NoStop}%
\bibitem [{sup()}]{supplx_deriv}%
  \BibitemOpen
  \href@noop {} {\bibinfo  {journal} {See supplementary material at
  .................. for detailed derivations}\ }\BibitemShut {NoStop}%
\bibitem [{\citenamefont {Chikazumi}(1964)}]{RefWorks:157}%
  \BibitemOpen
\bibfield  {journal} {  }\bibfield  {author} {\bibinfo {author} {\bibfnamefont
  {S.}~\bibnamefont {Chikazumi}},\ }\href@noop {} {\emph {\bibinfo {title}
  {{Physics of Magnetism}}}}\ (\bibinfo  {publisher} {Wiley New York},\
  \bibinfo {year} {1964})\BibitemShut {NoStop}%
\bibitem [{\citenamefont {Beleggia}\ \emph {et~al.}(2005)\citenamefont
  {Beleggia}, \citenamefont {Graef}, \citenamefont {Millev}, \citenamefont
  {Goode},\ and\ \citenamefont {Rowlands}}]{RefWorks:402}%
  \BibitemOpen
  \bibfield  {author} {\bibinfo {author} {\bibfnamefont {M.}~\bibnamefont
  {Beleggia}}, \bibinfo {author} {\bibfnamefont {M.~D.}\ \bibnamefont {Graef}},
  \bibinfo {author} {\bibfnamefont {Y.~T.}\ \bibnamefont {Millev}}, \bibinfo
  {author} {\bibfnamefont {D.~A.}\ \bibnamefont {Goode}}, \ and\ \bibinfo
  {author} {\bibfnamefont {G.~E.}\ \bibnamefont {Rowlands}},\ }\href@noop {}
  {\bibfield  {journal} {\bibinfo  {journal} {J. Phys. D: Appl. Phys.}\
  }\textbf {\bibinfo {volume} {38}},\ \bibinfo {pages} {3333} (\bibinfo {year}
  {2005})}\BibitemShut {NoStop}%
\bibitem [{\citenamefont {Brown}, \citenamefont {Novotny},\ and\ \citenamefont
  {Rikvold}(2001)}]{RefWorks:388}%
  \BibitemOpen
  \bibfield  {author} {\bibinfo {author} {\bibfnamefont {G.}~\bibnamefont
  {Brown}}, \bibinfo {author} {\bibfnamefont {M.~A.}\ \bibnamefont {Novotny}},
  \ and\ \bibinfo {author} {\bibfnamefont {P.~A.}\ \bibnamefont {Rikvold}},\
  }\href@noop {} {\bibfield  {journal} {\bibinfo  {journal} {Phys. Rev. B}\
  }\textbf {\bibinfo {volume} {64}},\ \bibinfo {pages} {134422} (\bibinfo
  {year} {2001})}\BibitemShut {NoStop}%
\bibitem [{\citenamefont {Abbundi}\ and\ \citenamefont
  {Clark}(1977)}]{RefWorks:179}%
  \BibitemOpen
  \bibfield  {author} {\bibinfo {author} {\bibfnamefont {R.}~\bibnamefont
  {Abbundi}}\ and\ \bibinfo {author} {\bibfnamefont {A.~E.}\ \bibnamefont
  {Clark}},\ }\href@noop {} {\bibfield  {journal} {\bibinfo  {journal} {IEEE
  Trans. Magn.}\ }\textbf {\bibinfo {volume} {13}},\ \bibinfo {pages} {1519}
  (\bibinfo {year} {1977})}\BibitemShut {NoStop}%
\bibitem [{\citenamefont {Ried}\ \emph {et~al.}(1998)\citenamefont {Ried},
  \citenamefont {Schnell}, \citenamefont {Schatz}, \citenamefont {Hirscher},
  \citenamefont {Ludescher}, \citenamefont {Sigle},\ and\ \citenamefont
  {Kronm\"{u}ller}}]{RefWorks:176}%
  \BibitemOpen
  \bibfield  {author} {\bibinfo {author} {\bibfnamefont {K.}~\bibnamefont
  {Ried}}, \bibinfo {author} {\bibfnamefont {M.}~\bibnamefont {Schnell}},
  \bibinfo {author} {\bibfnamefont {F.}~\bibnamefont {Schatz}}, \bibinfo
  {author} {\bibfnamefont {M.}~\bibnamefont {Hirscher}}, \bibinfo {author}
  {\bibfnamefont {B.}~\bibnamefont {Ludescher}}, \bibinfo {author}
  {\bibfnamefont {W.}~\bibnamefont {Sigle}}, \ and\ \bibinfo {author}
  {\bibfnamefont {H.}~\bibnamefont {Kronm\"{u}ller}},\ }\href@noop {}
  {\bibfield  {journal} {\bibinfo  {journal} {Phys. Stat. Sol. (a)}\ }\textbf
  {\bibinfo {volume} {167}},\ \bibinfo {pages} {195} (\bibinfo {year}
  {1998})}\BibitemShut {NoStop}%
\bibitem [{\citenamefont {Kellogg}\ and\ \citenamefont
  {Flatau}(2007)}]{RefWorks:178}%
  \BibitemOpen
  \bibfield  {author} {\bibinfo {author} {\bibfnamefont {R.}~\bibnamefont
  {Kellogg}}\ and\ \bibinfo {author} {\bibfnamefont {A.}~\bibnamefont
  {Flatau}},\ }\href@noop {} {\bibfield  {journal} {\bibinfo  {journal} {J.
  Intell. Mater. Sys. Struc.}\ }\textbf {\bibinfo {volume} {19}},\ \bibinfo
  {pages} {583} (\bibinfo {year} {2007})}\BibitemShut {NoStop}%
\bibitem [{\citenamefont {Cowburn}\ \emph {et~al.}(1999)\citenamefont
  {Cowburn}, \citenamefont {Koltsov}, \citenamefont {Adeyeye}, \citenamefont
  {Welland},\ and\ \citenamefont {Tricker}}]{RefWorks:133}%
  \BibitemOpen
  \bibfield  {author} {\bibinfo {author} {\bibfnamefont {R.~P.}\ \bibnamefont
  {Cowburn}}, \bibinfo {author} {\bibfnamefont {D.~K.}\ \bibnamefont
  {Koltsov}}, \bibinfo {author} {\bibfnamefont {A.~O.}\ \bibnamefont
  {Adeyeye}}, \bibinfo {author} {\bibfnamefont {M.~E.}\ \bibnamefont
  {Welland}}, \ and\ \bibinfo {author} {\bibfnamefont {D.~M.}\ \bibnamefont
  {Tricker}},\ }\href@noop {} {\bibfield  {journal} {\bibinfo  {journal} {Phys.
  Rev. Lett.}\ }\textbf {\bibinfo {volume} {83}},\ \bibinfo {pages} {1042}
  (\bibinfo {year} {1999})}\BibitemShut {NoStop}%
\bibitem [{\citenamefont {Jia}\ \emph {et~al.}(2006)\citenamefont {Jia},
  \citenamefont {Or}, \citenamefont {Chan}, \citenamefont {Zhao},\ and\
  \citenamefont {Luo}}]{pmnpt}%
  \BibitemOpen
  \bibfield  {author} {\bibinfo {author} {\bibfnamefont {Y.}~\bibnamefont
  {Jia}}, \bibinfo {author} {\bibfnamefont {S.~W.}\ \bibnamefont {Or}},
  \bibinfo {author} {\bibfnamefont {H.~L.~W.}\ \bibnamefont {Chan}}, \bibinfo
  {author} {\bibfnamefont {X.}~\bibnamefont {Zhao}}, \ and\ \bibinfo {author}
  {\bibfnamefont {H.}~\bibnamefont {Luo}},\ }\href@noop {} {\bibfield
  {journal} {\bibinfo  {journal} {Appl. Phys. Lett.}\ }\textbf {\bibinfo
  {volume} {88}},\ \bibinfo {pages} {242902} (\bibinfo {year}
  {2006})}\BibitemShut {NoStop}%
\bibitem [{\citenamefont {Lisca}\ \emph {et~al.}(2006)\citenamefont {Lisca},
  \citenamefont {Pintilie}, \citenamefont {Alexe},\ and\ \citenamefont
  {Teodorescu}}]{RefWorks:170}%
  \BibitemOpen
  \bibfield  {author} {\bibinfo {author} {\bibfnamefont {M.}~\bibnamefont
  {Lisca}}, \bibinfo {author} {\bibfnamefont {L.}~\bibnamefont {Pintilie}},
  \bibinfo {author} {\bibfnamefont {M.}~\bibnamefont {Alexe}}, \ and\ \bibinfo
  {author} {\bibfnamefont {C.~M.}\ \bibnamefont {Teodorescu}},\ }\href@noop {}
  {\bibfield  {journal} {\bibinfo  {journal} {Appl. Surf. Sci.}\ }\textbf
  {\bibinfo {volume} {252}},\ \bibinfo {pages} {4549} (\bibinfo {year}
  {2006})}\BibitemShut {NoStop}%
\bibitem [{\citenamefont {Masys}\ \emph {et~al.}(2003)\citenamefont {Masys},
  \citenamefont {Ren}, \citenamefont {Yang},\ and\ \citenamefont
  {Mukherjee}}]{RefWorks:563}%
  \BibitemOpen
  \bibfield  {author} {\bibinfo {author} {\bibfnamefont {A.~J.}\ \bibnamefont
  {Masys}}, \bibinfo {author} {\bibfnamefont {W.}~\bibnamefont {Ren}}, \bibinfo
  {author} {\bibfnamefont {G.}~\bibnamefont {Yang}}, \ and\ \bibinfo {author}
  {\bibfnamefont {B.~K.}\ \bibnamefont {Mukherjee}},\ }\href@noop {} {\bibfield
   {journal} {\bibinfo  {journal} {J. Appl. Phys.}\ }\textbf {\bibinfo {volume}
  {94}},\ \bibinfo {pages} {1155} (\bibinfo {year} {2003})}\BibitemShut
  {NoStop}%
\bibitem [{\citenamefont {Julliere}(1975)}]{RefWorks:577}%
  \BibitemOpen
  \bibfield  {author} {\bibinfo {author} {\bibfnamefont {M.}~\bibnamefont
  {Julliere}},\ }\href@noop {} {\bibfield  {journal} {\bibinfo  {journal}
  {Phys. Lett. A}\ }\textbf {\bibinfo {volume} {54}},\ \bibinfo {pages} {225}
  (\bibinfo {year} {1975})}\BibitemShut {NoStop}%
\bibitem [{\citenamefont {Moodera}\ \emph {et~al.}(1995)\citenamefont
  {Moodera}, \citenamefont {Kinder}, \citenamefont {Wong},\ and\ \citenamefont
  {Meservey}}]{RefWorks:555}%
  \BibitemOpen
  \bibfield  {author} {\bibinfo {author} {\bibfnamefont {J.~S.}\ \bibnamefont
  {Moodera}}, \bibinfo {author} {\bibfnamefont {L.~R.}\ \bibnamefont {Kinder}},
  \bibinfo {author} {\bibfnamefont {T.~M.}\ \bibnamefont {Wong}}, \ and\
  \bibinfo {author} {\bibfnamefont {R.}~\bibnamefont {Meservey}},\ }\href@noop
  {} {\bibfield  {journal} {\bibinfo  {journal} {Phys. Rev. Lett.}\ }\textbf
  {\bibinfo {volume} {74}},\ \bibinfo {pages} {3273} (\bibinfo {year}
  {1995})}\BibitemShut {NoStop}%
\bibitem [{\citenamefont {Mathon}\ and\ \citenamefont
  {Umerski}(2001)}]{RefWorks:572}%
  \BibitemOpen
  \bibfield  {author} {\bibinfo {author} {\bibfnamefont {J.}~\bibnamefont
  {Mathon}}\ and\ \bibinfo {author} {\bibfnamefont {A.}~\bibnamefont
  {Umerski}},\ }\href@noop {} {\bibfield  {journal} {\bibinfo  {journal} {Phys.
  Rev. B}\ }\textbf {\bibinfo {volume} {63}},\ \bibinfo {pages} {220403}
  (\bibinfo {year} {2001})}\BibitemShut {NoStop}%
\bibitem [{\citenamefont {Butler}\ \emph {et~al.}(2001)\citenamefont {Butler},
  \citenamefont {Zhang}, \citenamefont {Schulthess},\ and\ \citenamefont
  {MacLaren}}]{RefWorks:76}%
  \BibitemOpen
  \bibfield  {author} {\bibinfo {author} {\bibfnamefont {W.~H.}\ \bibnamefont
  {Butler}}, \bibinfo {author} {\bibfnamefont {X.~G.}\ \bibnamefont {Zhang}},
  \bibinfo {author} {\bibfnamefont {T.~C.}\ \bibnamefont {Schulthess}}, \ and\
  \bibinfo {author} {\bibfnamefont {J.~M.}\ \bibnamefont {MacLaren}},\
  }\href@noop {} {\bibfield  {journal} {\bibinfo  {journal} {Phys. Rev. B}\
  }\textbf {\bibinfo {volume} {63}},\ \bibinfo {pages} {054416} (\bibinfo
  {year} {2001})}\BibitemShut {NoStop}%
\bibitem [{\citenamefont {Yuasa}\ \emph {et~al.}(2004)\citenamefont {Yuasa},
  \citenamefont {Nagahama}, \citenamefont {Fukushima}, \citenamefont {Suzuki},\
  and\ \citenamefont {Ando}}]{RefWorks:74}%
  \BibitemOpen
  \bibfield  {author} {\bibinfo {author} {\bibfnamefont {S.}~\bibnamefont
  {Yuasa}}, \bibinfo {author} {\bibfnamefont {T.}~\bibnamefont {Nagahama}},
  \bibinfo {author} {\bibfnamefont {A.}~\bibnamefont {Fukushima}}, \bibinfo
  {author} {\bibfnamefont {Y.}~\bibnamefont {Suzuki}}, \ and\ \bibinfo {author}
  {\bibfnamefont {K.}~\bibnamefont {Ando}},\ }\href@noop {} {\bibfield
  {journal} {\bibinfo  {journal} {Nature Mater.}\ }\textbf {\bibinfo {volume}
  {3}},\ \bibinfo {pages} {868} (\bibinfo {year} {2004})}\BibitemShut {NoStop}%
\bibitem [{\citenamefont {Parkin}\ \emph {et~al.}(2004)\citenamefont {Parkin},
  \citenamefont {Kaiser}, \citenamefont {Panchula}, \citenamefont {Rice},
  \citenamefont {Hughes}, \citenamefont {Samant},\ and\ \citenamefont
  {Yang}}]{RefWorks:33}%
  \BibitemOpen
  \bibfield  {author} {\bibinfo {author} {\bibfnamefont {S.~S.~P.}\
  \bibnamefont {Parkin}}, \bibinfo {author} {\bibfnamefont {C.}~\bibnamefont
  {Kaiser}}, \bibinfo {author} {\bibfnamefont {A.}~\bibnamefont {Panchula}},
  \bibinfo {author} {\bibfnamefont {P.~M.}\ \bibnamefont {Rice}}, \bibinfo
  {author} {\bibfnamefont {B.}~\bibnamefont {Hughes}}, \bibinfo {author}
  {\bibfnamefont {M.~G.}\ \bibnamefont {Samant}}, \ and\ \bibinfo {author}
  {\bibfnamefont {S.~H.}\ \bibnamefont {Yang}},\ }\href@noop {} {\bibfield
  {journal} {\bibinfo  {journal} {Nature Mater.}\ }\textbf {\bibinfo {volume}
  {3}},\ \bibinfo {pages} {862} (\bibinfo {year} {2004})}\BibitemShut {NoStop}%
\bibitem [{\citenamefont {Gallagher}\ and\ \citenamefont
  {Parkin}(2006)}]{RefWorks:300}%
  \BibitemOpen
  \bibfield  {author} {\bibinfo {author} {\bibfnamefont {W.~J.}\ \bibnamefont
  {Gallagher}}\ and\ \bibinfo {author} {\bibfnamefont {S.~S.~P.}\ \bibnamefont
  {Parkin}},\ }\href@noop {} {\bibfield  {journal} {\bibinfo  {journal} {IBM J.
  Res. Dev.}\ }\textbf {\bibinfo {volume} {50}},\ \bibinfo {pages} {5}
  (\bibinfo {year} {2006})}\BibitemShut {NoStop}%
\end{thebibliography}
%

\end{document}


\preprint{AIP/123-QED}

\title{Supplementary Information\\Critical analysis and remedy of switching failures in straintronic logic using Bennett clocking in the presence of thermal fluctuations}

\author{Kuntal Roy}
\email{royk@purdue.edu.}
\noaffiliation
\affiliation{School of Electrical and Computer Engineering, Purdue University, West Lafayette, Indiana 47907, USA}
\thanks{Work for this paper was performed prior to K. Roy joining Purdue University. Some affiliated work was performed at the School of Electrical and Computer Engineering, Virginia Commonwealth University, Richmond,  Virginia 23284, USA.}


\maketitle

\textit{Solution of stochastic Landau-Lifshitz-Gilbert (LLG) equation of magnetization dynamics in the presence of thermal fluctuations.--} We adopt the standard spherical coordinate system (see Fig.~1a in the main Letter) to solve the magnetization dynamics using stochastic Landau-Lifshitz-Gilbert (LLG) equation in the presence of room-temperature (300 K) thermal fluctuations.~\cite{roy11_6} The potential energy of the nanomagnet-2 can be expressed as the sum of the anisotropies due to shape and stress, and due to dipolar coupling from the nanomagnets 1 and 3,
\begin{align}
E_{total,2} &= E_{shape,2} + E_{stress,2} + E_{dipole,2} \nonumber\\
			&= B_2(\phi_2) \, sin^2 \theta_2 + E_{dipole,2},
\label{eq:total_anisotropy}
\end{align}
where
\begin{subequations}
\begin{align}
B_2(\phi_2) &= B_{shape,2}(\phi_2) + B_{stress,2},\displaybreak[3]\\
B_{shape,2}(\phi_2) &= (\mu_0/2) M_s^2 \Omega [(N_{d-yy}-N_{d-zz}) \nonumber \\ 
								& \qquad \qquad + (N_{d-xx}-N_{d-yy})\,cos^2\phi_2],\\
B_{stress,2} 	&= (3/2) \lambda_s \sigma_2 \Omega,
\label{eq:shape_stress}
\end{align}
\end{subequations}
\begin{align}
	E_{dipole,2} &= \cfrac{\mu_0}{4\pi R^3} \, M_s^2 \Omega^2 \lbrack cos\theta_2 cos\theta_1  + cos\theta_2 cos\theta_3 \nonumber \displaybreak[3]\\
	& \quad + sin \theta_1 sin \theta_2 (cos \phi_1 cos\phi_2 - 2 sin\phi_1 sin\phi_2) \nonumber \displaybreak[3]\\
	& \quad + sin \theta_3 sin \theta_2 (cos \phi_3 cos\phi_2 - 2 sin\phi_3 sin\phi_2) \rbrack \displaybreak[3],
\end{align}
$M_s$ is the saturation magnetization, $N_{d-mm}$ is the component of demagnetization factor along $m$-direction, which depends on the nanomagnet's dimensions,~\cite{RefWorks:157,RefWorks:402} $(3/2)\lambda_s$ is the magnetostrictive coefficient of the single-domain magnetostrictive nanomagnet,~\cite{RefWorks:157} $\sigma_2$ is the stress on the nanomagnet-2, and $R$ is the center-to-center distance between the nanomagnets. Note that the product of magnetostrictive coefficient and stress needs to be \emph{negative} in sign for stress-anisotropy to overcome the shape-anisotropy.

The magnetization $\mathbf{M_2}$ of the nanomagnet-2 has a constant magnitude  but a variable direction, so that we can represent it by a vector of unit norm $\mathbf{n_{m,2}} =\mathbf{M_2}/|\mathbf{M_2}| = \mathbf{\hat{e}_r}$ where $\mathbf{\hat{e}_r}$ is the unit vector in the radial direction in spherical coordinate system represented by ($r$,$\theta$,$\phi$). The effective field and torque acting on the magnetization due to gradient of potential landscape as in Eq.~\eqref{eq:total_anisotropy} can be expressed as $\mathbf{H_{eff,2}} = - \nabla E_{total,2} = - (\partial E_{total,2}/\partial \theta_2)\,\mathbf{\hat{e}_{\theta}} - (1/sin\theta_2)\,(\partial E_{total,2}/\partial \phi_2)\,\mathbf{\hat{e}_\phi}$ and $\mathbf{T_{E,2}}= \mathbf{n_{m,2}} \times \mathbf{H_{eff,2}}$, respectively. The thermal field and the corresponding torque acting on the magnetization can be written as $\mathbf{H_{TH,2}}=P_{\theta_2}\,\mathbf{\hat{e}_\theta}+P_{\phi_2}\,\mathbf{\hat{e}_\phi}$ and $\mathbf{T_{TH,2}}=\mathbf{n_{m,2}} \times \mathbf{H_{TH,2}}$, respectively,~\cite{RefWorks:186,roy11_6} where
\begin{subequations}
\begin{align}
P_{\theta_2} &= M_V [ h_{x,2}\,cos\theta_2\,cos\phi_2 + h_{y,2}\,cos\theta_2 sin\phi_2  \nonumber \displaybreak[3]\\
						 & \qquad \qquad- h_{z,2}\,sin\theta_2], \displaybreak[3]\\
P_{\phi_2} &= M_V [h_{y,2}\,cos\phi_2 -h_{x,2}\,sin\phi_2], \displaybreak[3]\\
h_{i,2} &= \sqrt{\frac{2 \alpha kT}{|\gamma| M_V \Delta t}} \;G_{(0,1)} \;\;(i=x,y,z), \label{eq:thermal_h}
\end{align}
\end{subequations}
$\alpha$ is the phenomenological damping parameter, $\gamma$ is the gyromagnetic ratio for electrons, $M_V= \mu_0 M_s \Omega$, $\Delta t$ is the simulation time-step, $G_{(0,1)}$ is a Gaussian distribution with zero mean and unit variance,~\cite{RefWorks:388} $k$ is the Boltzmann constant, and $T$ is temperature.

The magnetization dynamics under the two aforesaid torques is described by the stochastic Landau-Lifshitz-Gilbert (LLG) equation~\cite{RefWorks:162,RefWorks:161,RefWorks:186} as 
\begin{equation}
\cfrac{d\mathbf{n_{m,2}}}{dt} - \alpha \left(\mathbf{n_{m,2}} \times \cfrac{d\mathbf{n_{m,2}}}{dt} \right)\\
 = -\cfrac{|\gamma|}{M_V} \left\lbrack \mathbf{T_{E,2}} +  \mathbf{T_{TH,2}}\right\rbrack.
\end{equation}
After solving the LLG equation, we get the following coupled equations for the dynamics of $\theta_2$ and $\phi_2$:
\begin{multline}
\left(1+\alpha^2 \right) \cfrac{d\theta_2}{dt} = \cfrac{|\gamma|}{M_V} [ B_{shape,\phi_2}(\phi_2)sin\theta_2 \\
 - 2\alpha B_2(\phi_2) sin\theta_2 cos\theta_2 - T_{dipole,\theta_2} - \alpha T_{dipole,\phi_2} \\
 + (\alpha P_{\theta_2} + P_{\phi_2})],
 \label{eq:theta_dynamics_bennett}
\end{multline}
\begin{multline}
\left(1+\alpha^2 \right) \cfrac{d\phi_2}{dt} = \cfrac{|\gamma|}{M_V} \cfrac{1}{sin\theta_2} [\alpha B_{shape,\phi_2}(\phi_2)sin\theta_2 \\
	 + 2 B_2(\phi_2) sin\theta_2 cos\theta_2 + \alpha T_{dipole,\theta_2} + T_{dipole,\phi_2} \\
	 - \{sin\theta_2\}^{-1} (P_{\theta_2} - \alpha P_{\phi_2})] \qquad (sin\theta_2 \neq 0),
  \label{eq:phi_dynamics_bennett}
\end{multline}
where
\begin{align}
B_{shape,\phi_2}(\phi_2) &= -\cfrac{\partial B_{shape,2}(\phi_2)}{\partial \phi_2} \nonumber\\
												 &= (\mu_0/2) \, M_s^2 \Omega (N_{d-xx}-N_{d-yy}) sin(2\phi_2),
\label{eq:B_shape_phi}
\end{align}
$T_{dipole,\theta_2} = (1/sin \theta_2)(\partial E_{dipole,2}/\partial \phi_2)$, and $T_{dipole,\phi_2} = \partial E_{dipole,2}/\partial \theta_2$. Note that in a very similar way the equations of dynamics for the other three nanomagnets can be derived.

The energy dissipation in the nanomagnet-2 due to Gilbert damping can be expressed as $E_{d,2}=\int_0^\tau P_{d,2}(t)\,dt$, where $\tau$ is the switching delay and the instantaneous power dissipation can be calculated as
\begin{equation}
P_{d,2}(t) = \cfrac{\alpha \, |\gamma|}{(1+\alpha^2) M_V} \, \left| \mathbf{T_{E,2}}(t)\right|^2.
\label{eq:power_dissipation_dipole}
\end{equation}
While calculating internal energy dissipation, we sum up the energy dissipations in all the four nanomagnets but note that the dissipations in the nanomagnet-1 and nanomagnet-4 are quite negligible since they don't rotate much and the dissipation in nanomagnet-3 is about half of that in nanomagnet-2 (since nanomagnet-2 switches a complete $180^\circ$, while the nanomagnet-3 switches only about $90^\circ$). We sum up this internal energy dissipation with the energy dissipation in the external circuitry (which is miniscule~\cite{roy11_2,roy11_6}) to get the total energy dissipation.

The performance metrics switching delay and energy dissipation are determined by solving stochastic Landau-Lifshitz-Gilbert equation in the presence of room-temperature (300 K) thermal fluctuations. We assume a 20 MPa of stress and we reverse the stress during ramp-down phase to aid switching.~\cite{roy11_6} We perform a moderately large number (10000) of simulations and we consider that magnetization's initial orientation is \emph{not} fixed rather it is fluctuating due to thermal agitations, so we have a distribution of magnetization's initial orientation when magnetization starts switching. We determine the initial distributions of both polar angle $\theta$ and azimuthal angle $\phi$ at room-temperature by solving the stochastic LLG equation when no stress is active.~\cite{roy13_2,roy11_6} When magnetization of nanomagnet-2 reaches $\theta \leq 5^\circ$, the switching is deemed to have completed.

\makeatletter 
\renewcommand\@biblabel[1]{$^{S#1}$}
\makeatother

%